\documentclass[12pt]{article}
\usepackage{times}
\usepackage{geometry}
\usepackage[utf8]{inputenc}
\usepackage{enumitem,amssymb}
\usepackage{ragged2e}
\newlist{thematic}{itemize}{8}
\setlist[thematic]{label=$\square$}
\usepackage{pifont}
\usepackage{ulem}
\usepackage{jheppub}   
\usepackage[dvipsnames,svgnames,x11names]{xcolor}

\usepackage{sectsty}
\sectionfont{\fontsize{15}{15}\selectfont}

\usepackage{sidecap}
\sidecaptionvpos{figure}{c}

\definecolor{DarkGreen}{rgb}{0.0, 0.3, 0.0}
\definecolor{purple}{rgb}{0.5, 0.0, 0.5}
\definecolor{red}{rgb}{1, 0.0, 0.0}
\definecolor{green}{rgb}{0, 1.0, 0.0}

\def\3he{$^3{\rm He}$}

\def\lsim{\mathrel{\lower2.5pt\vbox{\lineskip=0pt\baselineskip=0pt
           \hbox{$<$}\hbox{$\sim$}}}}

\def\gsim{\mathrel{\lower2.5pt\vbox{\lineskip=0pt\baselineskip=0pt
           \hbox{$>$}\hbox{$\sim$}}}}

\begin{document}
\raggedright
\Large
\textbf{Astro2020 Science White Paper} \linebreak

\LARGE
``SZ spectroscopy" in the coming decade: Galaxy cluster cosmology and astrophysics in the submillimeter \linebreak
\normalsize


\noindent \textbf{Thematic Areas:} 
\vspace{-2mm}
\begin{flushleft}
{\normalfont
  \large \textsc{Primary:} \textit{Cosmology and Fundamental Physics} \\
\textsc{Secondary:} \textit{Galaxy Evolution} \\}
\end{flushleft}
\medskip

\textbf{Corresponding Author:}

Name: Kaustuv Basu	
 \linebreak						
Institution: University of Bonn 
 \linebreak
Email: kbasu@astro.uni-bonn.de
 \linebreak
Phone: $+$49 228 735 658 
 \linebreak
  
\vspace{2mm}
\begin{minipage}[s]{0.95\columnwidth}

\textbf{Co-authors:} 
Jens Erler (Bonn), 
Jens Chluba (Manchester), 
Jacques Delabrouille (APC Paris), 
J.~Colin Hill (IAS/Flatiron Institute), 
Tony Mroczkowski (ESO), 
Michael D. Niemack (Cornell), 
Mathieu Remazeilles (Manchester), 
Jack Sayers (Caltech), 
Douglas Scott (UBC), 
Eve M. Vavagiakis (Cornell), 
Michael Zemcov (RIT),   
Manuel Aravena (UDP Santiago), 
James G. Bartlett (APC Paris/JPL), 
Nicholas Battaglia (Cornell), 
Frank Bertoldi (Bonn), 
Maude Charmetant (Bonn), 
Sunil Golwala (Caltech), 
Terry L. Herter (Cornell), 
Pamela Klaassen (UK ATC), 
Eiichiro Komatsu (MPA), 
Benjamin Magnelli (Bonn), 
Adam B. Mantz (KIPAC/Stanford), 
P.~Daniel Meerburg (KICC/Groningen), 
Jean-Baptiste Melin (IRFU Saclay), 
Daisuke Nagai (Yale), 
Stephen C. Parshley (Cornell), 
Etienne Pointecouteau (IRAP Toulouse), 
Miriam E. Ramos-Ceja (Bonn), 
Mateusz Ruszkowski (Michigan), 
Neelima Sehgal (Stony Brook), 
Gordon G. Stacey (Cornell), 
Rashid Sunyaev (MPA/IKI) 
  \linebreak 
  
\end{minipage}
  


\justify 
\textbf{Abstract:} 
Sunyaev-Zeldovich (SZ) effects were first proposed in the 1970s as tools to identify the X-ray emitting hot gas inside massive clusters of galaxies and obtain their velocities relative to the cosmic microwave background (CMB).  
Yet it is only within the last decade that they have begun to significantly impact astronomical research. 
Thanks to the rapid developments in CMB instrumentation, measurement of the dominant thermal signature of the SZ effects has become a routine tool to find and characterize large samples of galaxy clusters and to seek deeper understanding of several important astrophysical processes via high-resolution imaging studies of many targets. With the notable exception of the {\sl Planck} satellite and a few combinations of ground-based observatories, much of this ``SZ revolution'' has happened in the photometric mode, where observations are made at one or two frequencies in the millimeter regime to maximize the cluster detection significance and minimize the foregrounds. Still, there is much more to learn from detailed and systematic analyses of the SZ spectra across multiple wavelengths, specifically in the submillimeter ($\gtrsim$ 300~GHz) domain. 
The goal of this Science White Paper is to highlight this particular aspect of SZ research, point out what new and potentially groundbreaking insights can be obtained from these studies, and emphasize why the coming decade can be a golden era for SZ spectral measurements.

\pagebreak
\section{Introduction to the SZ landscape}
\vspace{-2mm}

Galaxy clusters stand at the crossroads between astrophysics and cosmology. By forming the most massive end of the dark-matter halo mass function, their number counts deliver effective constraints on the composition and growth history of the Universe \cite{Allen11,Kravtsov12,Nagai14}. At the same time, galaxy clusters provide unique laboratories to test several astrophysical phenomena -- from massive galaxy evolution and the role of AGN feedback, to particle acceleration in Mpc-scale shocks \cite{Sarazin86,Kormendy89,Bykov00}. Quite naturally, studies of galaxy clusters and the associated large-scale structures have been one of the most productive areas of research in the last few decades, collecting huge amounts of data from ground and space based observatories across the entire electromagnetic spectrum.

Among the various methods to find and characterize galaxy clusters, one of the newest and most rapidly developing is the Sunyaev-Zeldovich effect \cite{SZ70,SZ72,SZ75,SZ80rev,Birkinshaw99,Carlstrom02,SSRev2019}. It has two main variants: the thermal (tSZ) and the kinematic (kSZ) effects. 
They arise from inverse Compton scattering of the CMB photons by hot intracluster electrons and have several desirable properties: the signals are practically redshift independent (only limited by the telescope beam), have unique spectral signatures, and the 
amplitudes connect directly to total cluster thermal energy and line-of-sight momenta. 
From the first blind tSZ detection of clusters only a decade ago \cite{Stanis09,Menanteau10}, catalogs now exist with over a thousand confirmed objects and an order-of-magnitude more are expected  from the next-generation CMB experiments \cite{SOscience,CMB-S4,COREclust},  
 transforming cosmological quests such as the nature of dark energy and neutrino masses. 
Another rapidly developing field is the high angular resolution SZ imaging \cite{SSRev2019} that is opening up new windows on  cluster astrophysics 
(see white papers by Mroczkowski et al. and Sehgal et al.).
Our aim in this paper is to highlight some of the unique science that can be addressed from detailed measurements of the SZ spectra  and how those results can shape our view of the Universe within the decade 2020--2030 and beyond.

\begin{figure}[tbh]
\begin{center}
  \includegraphics[width=\textwidth, height=6cm, clip=true, trim=0 170 0 150]{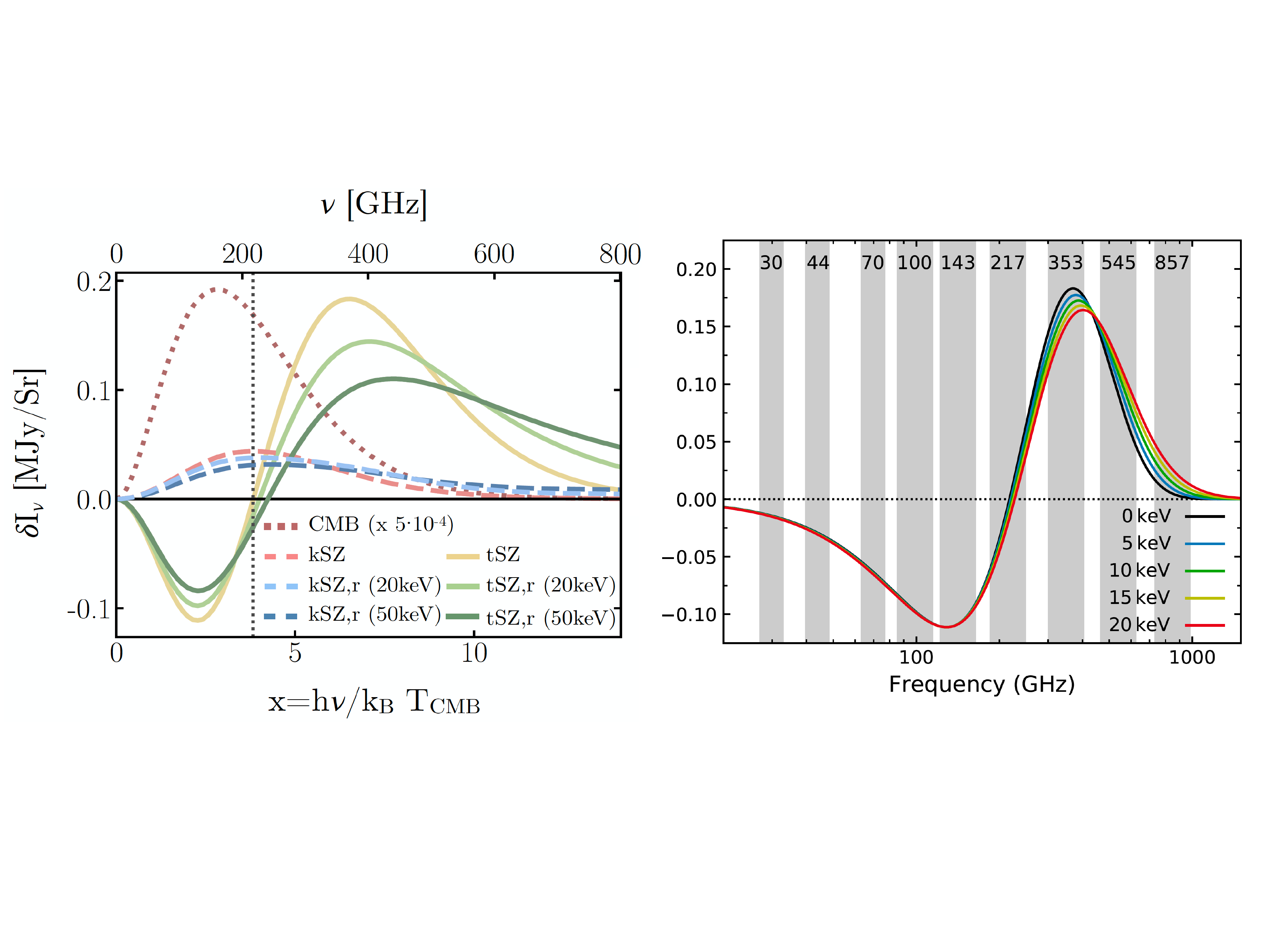}
\end{center}
\vspace{-10mm}
\caption{\small 
{\it Left:} tSZ (solid) and kSZ (dashed) spectra including relativistic corrections from high-energy thermal electrons. Two extreme temperatures highlight the spectral shapes, also for the kinematic effect. The dotted curve is the unscattered CMB scaled down by a factor of 2000 (figure from Ref.~\cite{SSRev2019}). {\it Right:} tSZ spectral distortions once again, with a more representative set of temperatures and a varying Compton-$y$ that cause the spectra to match at the minima of the tSZ decrement. 
This panel illustrates the importance of high frequency observations to break the $y-T_e$ degeneracy. 
The vertical grey bands are {\sl Planck} frequency channels, many of which are also accessible from a good ground site, e.g., in Atacama or Antarctica.
}
\label{fig:spectra}
\end{figure}


The importance of submillimeter data to separate the kSZ effect was realized early on \cite{Lamarre98,Komatsu99}.  
This effect carries information about the electron bulk motion relative to the CMB rest frame and promises to be an important cosmological probe \cite{DeDeo05,Bhatta08,Mueller15,Schaan16,Sugiyama17}, 
yet successful applications have been limited, most notably for identifying strong internal bulk motions in merging clusters \cite{Mroczkowski12,Sayers13,Adam17} and detecting the kSZ-galaxy bispectrum via pairwise subtraction  \cite{Hand12,PlXIII14,Soergel18}. Terrestrial CMB experiments have mostly used the ``photometric mode" for SZ science: focusing on 1--2 bands in the millimeter to maximize tSZ detection significance and using the 220 GHz channel to subtract the CMB, which also eliminates the kSZ signal.  
This has kept the focus on a {\it static} view of the universe to perform number count cosmology similar to X-ray or infra-red surveys. 
Now, after the {\sl Planck} data release, the real potential of multi-frequency SZ observations is coming into perspective, from  kSZ measurements and the search for the missing baryons \cite{PlXXXVII16,Hill16,Lim17}, to extracting cluster properties from the tSZ spectrum \cite{Hurier17,Erler2018}.
The next decade will see a full realization of these scientific quests when large format ground-based cameras will improve upon the resolution and sensitivity of past satellite probes, including in the submm, while at the same time new space missions will offer unprecedented spectral sampling and sky coverage to complement the results from the ground.

\vspace{-2mm}
\section{Galaxy cluster temperatures from the rSZ effect}
\vspace{-2mm}
\textit{What will change if we can measure galaxy cluster temperatures across cosmic time?} \\
\vspace{-6mm}

\begin{figure}[tbh]
\begin{center}
  \includegraphics[width=\textwidth, height=6.5cm, clip=true, trim=10 150 0 170]{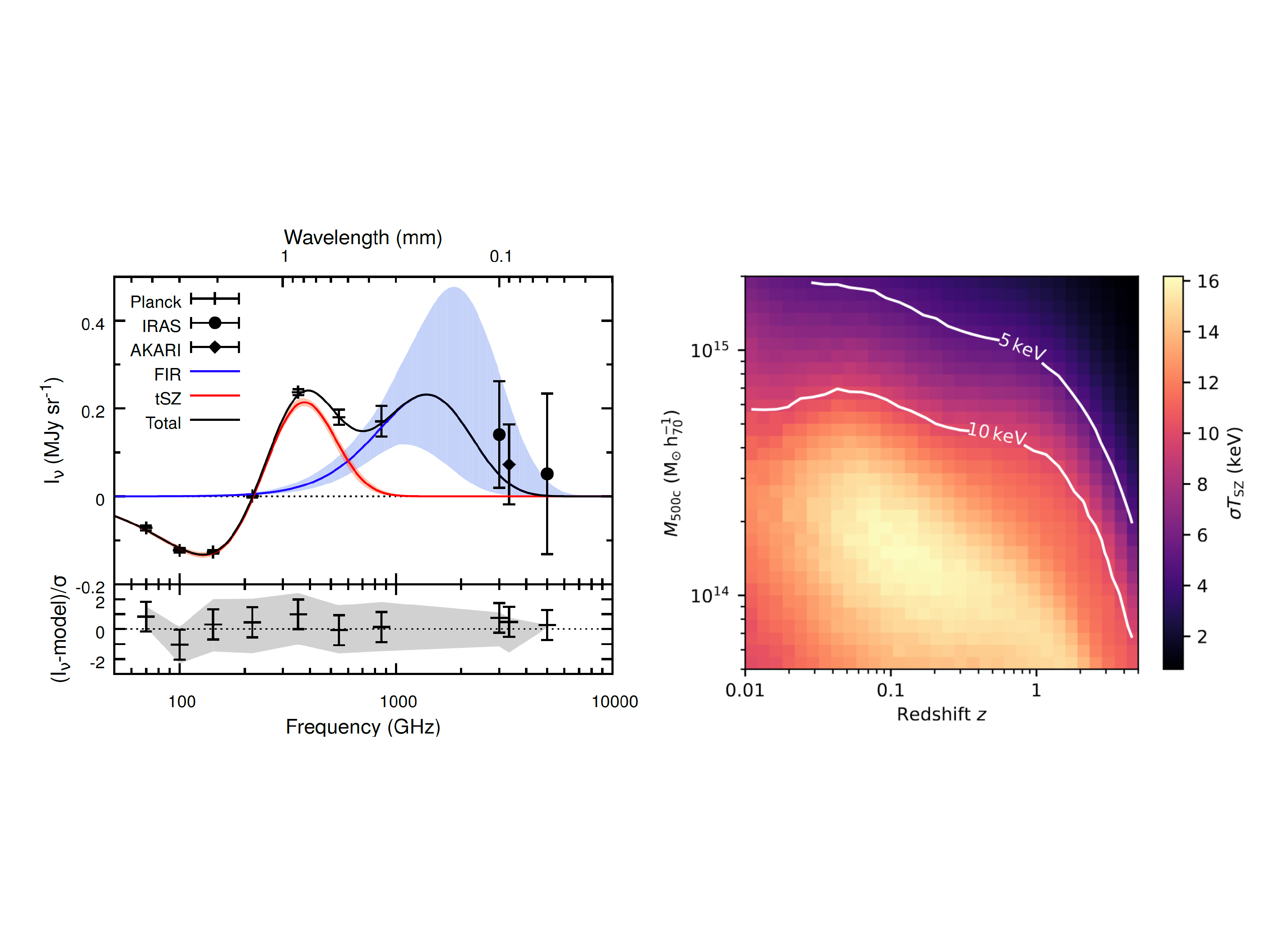}
\end{center}
\vspace{-10mm}
\caption{\small 
{\it Left:} Spectral modeling result from a stacked sample of 772 {\sl Planck} clusters, with red and blue lines denoting the best-fit tSZ and thermal dust emission spectra. The mean cluster temperature is measured only at $2.2\sigma$ significance ($4.4^{+2.1}_{-2.0}$ keV), but the result shows the importance of accurate dust modeling using high-frequency data (figure from Ref.~\cite{Erler2018}). {\it Right:} rSZ temperature measurement uncertainties on individual clusters as expected from an ``optimized CCAT-prime" camera concept \cite{Mittal2018} with 10,000~deg$^2$ survey area. The errors actually improve with redshift, in complete contrast with X-ray observations, enabling us to stack and calibrate cluster masses out to the epoch of their formation. Furthermore, future surveys can improve these predictions by employing additional frequency bands or using priors from X-ray data. 
}
\label{fig:rsz}
\end{figure}
\vspace{-2mm}

Just as with the X-ray Bremsstrahlung spectrum, the full, relativistic spectrum of the tSZ effect (referred also as the relativistic SZ, or rSZ, effect) carries information about the mean temperature of the scattering electrons \cite{Wright79,Pointecouteau98,Nozawa09,Chluba12}. As shown in Fig.~\ref{fig:spectra} (right panel), mm-wave observations alone cannot distinguish this effect from a change in the Comptonization parameter, and only by adding high-frequency observations it is possible to break this $y-T_e$ degeneracy and extract the temperature information 
(unless other priors on electron densities are used). 
The current best constraints on the tSZ spectrum and the associated $T_\mathrm{rSZ}$ value comes from {\sl Planck} satellite data (Fig.~\ref{fig:rsz} left), which provide a marginal detection after stacking hundreds of clusters \cite{Hurier16,Erler2018}. But next-generation CMB experiments with submm capabilities will push the noise down to levels where similar detection significance can be achieved on {\it individual} massive clusters \cite{Mittal2018,Erler2018}. 
When measured via stacking (which will eliminate the kSZ contribution) 
the accuracy of $T_\mathrm{rSZ}$-based mass calibration will be similar to what is expected from CMB lensing and other methods, hence providing an important tool to model the thermodynamic history of galaxy clusters from a very early epoch. 

An example is shown in Fig.~\ref{fig:rsz} (right), where $T_\mathrm{rSZ}$ errors are computed for an SZ survey with three submm bands \cite{Mittal2018} and realistic foregrounds, yielding temperature accuracies above $6\sigma$ in stacked cluster samples at $z\sim1$. 
This will be a significant step forward, since the temperature determination from forthcoming X-ray surveys such as SRG/eROSITA will be limited mostly to $z\;\lesssim\;0.2$ clusters \cite{Borm14,Hofmann17}. $T_\mathrm{rSZ}$ is further interesting as it is pressure ($n_eT_e$) weighted \cite{Hansen04,Kay08}, as opposed to roughly $n_e^2$ weighting of the X-ray temperature \cite{Mazzotta04}, hence it is a low-scatter mass proxy suitable for cosmological modeling of the halo mass function. Accounting for the relativistic corrections will also improve the accuracy of cosmological modeling from tSZ power spectra and possibly alleviate some of the current tensions between CMB and cluster cosmology results \cite{Remazeilles2019}. 

{\sl Planck} satellite data in the present decade have also provided unmistakable signs of thermal emission from dust within galaxy clusters themselves \cite{PlDust1,PlDust2,Melin2018}, which will be a critical component for SZ spectral modeling. 
The origin of this cluster-centric FIR emission is currently unknown; it could be from individual star-forming galaxies (i.e. a component of the cosmic infrared background) but could also be diffuse dust,  
accumulated from galaxy stripping \cite{Vogelsberger18} or AGN-uplifting of the central cold gas \cite{Werner2010}. 
Its impact will be even more significant for high-$z$ proto-clusters or for the study of the circumgalactic medium (CGM) in low-mass halos, where the relative dust contribution could be higher. It has already been shown that, for modeling AGN feedback in galaxy halos from the associated tSZ signal, not accounting for the dust emission can lead to incorrect conclusions \cite{Soergel2017}. If our goal is to calibrate cluster masses across cosmic time using the rSZ effect, then accurate modeling of their dust emission using submm data is going to play a central role.

\vspace{-2mm}
\section{SZ component separation and the measurement of cluster velocities}
\vspace{-2mm}
\textit{How can we ensure unbiased kSZ measurements and what will be the scientific impact?} \\
\vspace{-3mm}

\begin{figure}[tbh]
\begin{center}
  \includegraphics[width=\textwidth, height=7cm, clip=true, trim=0 140 0 165]{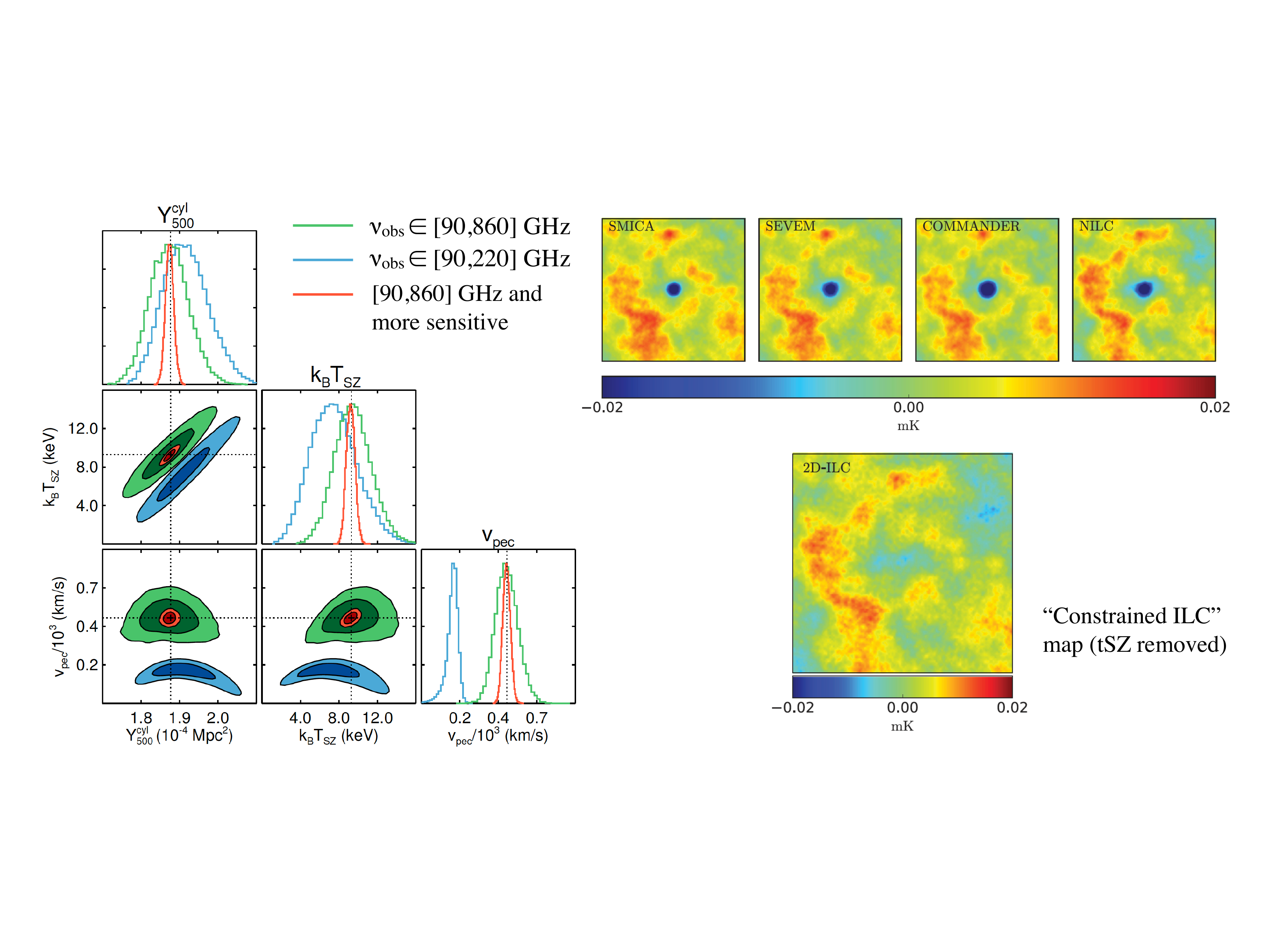}
\end{center}
\vspace{-8mm}
\caption{\small 
{\it Left:} Measurement forecasts for the basic SZ parameters 
in a massive galaxy cluster harboring dust, using only the thermal noise. One instrument is limited to the millimeter wavebands (blue) and does not include dust in the modeling, giving biased estimates for the temperature and velocity. Instruments that extend into the submm (green and red) yield unbiased results after marginalizing over dust  
(see Ref.~\cite{Stacey2018}). {\it Right:} Residuals from the tSZ signal in the direction of known clusters as seen in the public-release {\sl Planck} 2015 CMB maps (top row), and one map where this residual is minimized via specialized techniques. When not corrected,  this residual will cause a bias in any kind of cross-correlation study between the CMB/kSZ and large-scale structure probes, and similar residuals can be expected from cluster dust 
(figures from Ref.~\cite{Chen2018}).
}
\vspace{-3mm}
\label{fig:bias}
\end{figure}

\noindent The issue of dust contamination will enter the study of the kSZ effect -- and hence the determination of the cosmic velocity field on large scales -- in two ways. First, similar to the rSZ temperature measurement, it will bias the line-of-sight velocity estimation for individual halos. This problem is illustrated in the left panel of Fig.~\ref{fig:bias}, where an instrument that has only low-frequency (mm-wave) coverage returns biased estimates for temperature and velocities by ignoring dust emission, while having no discernible bias on the Comptonization parameter. 
Even though statistically not the most powerful technique, measuring the velocity (and hence optical depth) of selected individual  high-mass halos will be extremely important for calibrating the galaxy-electron density power spectrum and breaking the so-called ``optical depth degeneracy" \cite{Battaglia16,Flender17}. Such measurements will also be important in the search for missing baryons from stacked electron density profiles \cite{PlXXXVII16,Lim17}.

The second case of dust contamination will be within CMB maps themselves, which are the templates for kSZ signal in every kSZ bispectrum or cross-correlation analyses (at small angular scales the primordial CMB power is mostly replaced by kSZ). 
The origin of this bias is incorrect/insufficient use of foreground information in map making.  
An illustrative example is shown in Fig.~\ref{fig:bias} (right), where all of the four publicly-released {\sl Planck} CMB maps from 2015 show strong residuals in the direction of known clusters. This residual comes from not explicitly accounting for the tSZ signal in the foreground model and minimizing that contribution \cite{Chen2018}. For the high-precision CMB imaging in the coming decade, a simple non-relativistic tSZ template will not suffice, but one will need its relativistic corrections as well \cite{Remazeilles2019}, necessitating submm data. The same will be true with dust, whose correlation with large-scale structure is already proven \cite{PlDust1}, thereby also requiring the leverage of high-frequency data for unbiased CMB map extraction. 

The impact of kSZ measurements on cosmology, via both direct and statistical methods, will be immense (see white paper by Battaglia et al.). It will help to identify the time evolution of dark energy from large-scale velocity correlations \cite{Alonso16}, search for missing baryons at low redshifts \cite{Hernandez08}, and provide constraints on the energy feedback within galaxy halos \cite{Battaglia17}, to list a few. High-frequency ($\gtrsim$ 300~GHz) CMB observations will be a critical ingredient for building this dynamical view of the Universe, together with the next-generation infra-red, optical, and X-ray surveys.

\vspace{-2mm}
\section{ntSZ effect for the cosmic ray energy budget of clusters}
\vspace{-2mm}
\textit{What role does the nonthermal population play in determining large-scale structure growth?} \\
\vspace{-3mm}

\begin{SCfigure}[1.4][h!]
  \hspace{-4mm}
   \includegraphics[width=7.5cm, height=5cm, clip=true, trim=165 115 165 140]{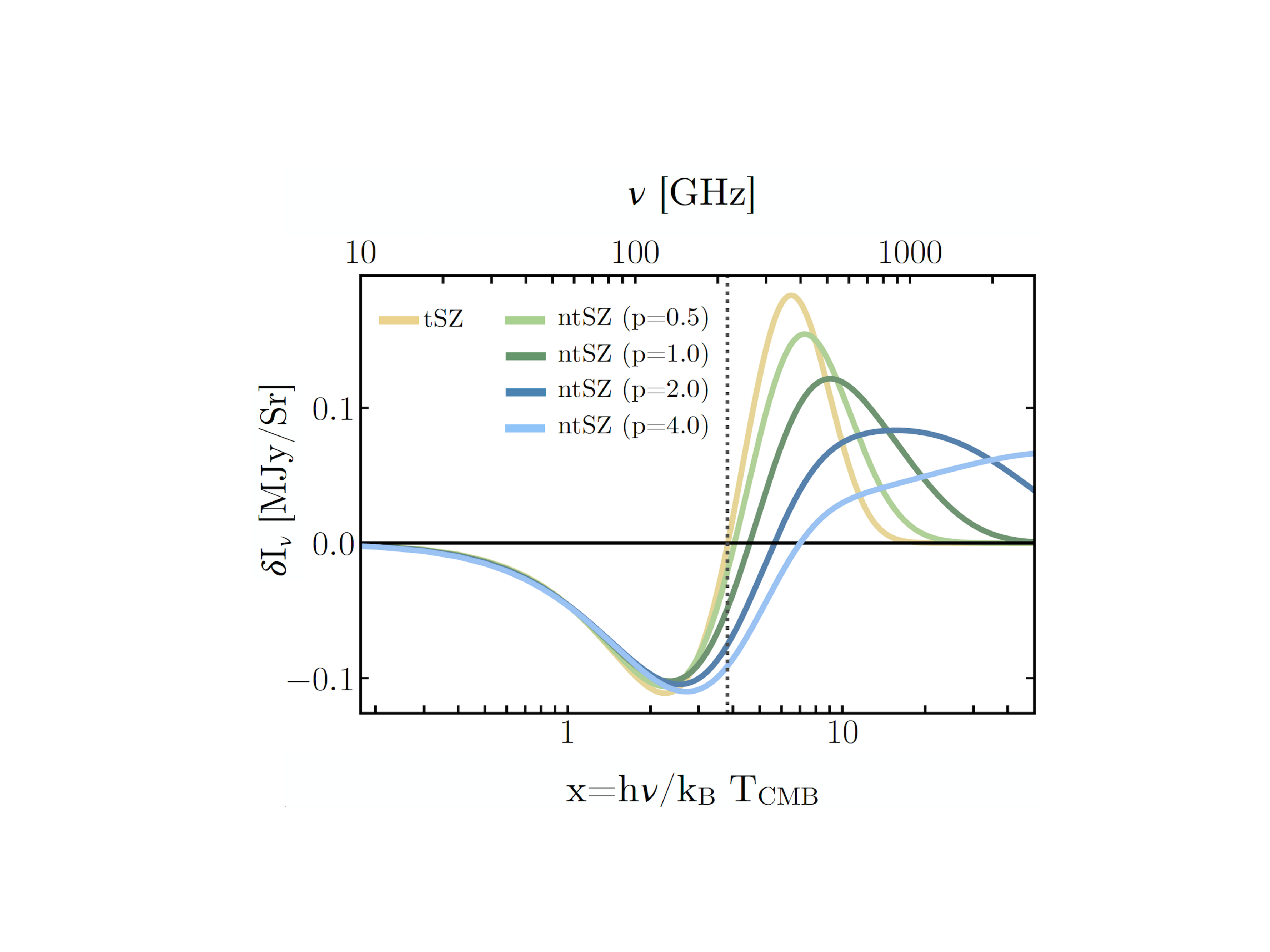}
   \caption{\small Comparison between the non-relativistic tSZ spectrum and several non-thermal spectra with mono-energetic electrons (denoted by their dimensionless momentum $p$, where $p=\sqrt{\gamma^2 - 1}$ in the relativistic limit). In reality the ntSZ spectra are more complex, due to the power-law energy distribution of non-thermal particles as well as for multiple power law indices within clusters; but this plot highlights the importance of submm observations to disentangle the small effect (figure from Ref.~\cite{SSRev2019}).
   }\label{fig:ntsz}
\end{SCfigure}

\noindent Exploration of the SZ spectral distortions will not be complete without taking into account the non-thermal SZ (ntSZ) effect \cite{Rephaeli95,Enss00,Colafrancesco03}. As shown in Fig.~\ref{fig:ntsz}, the high-frequency bands are again important in disentangling its contribution, but the spectral shapes are more uncertain due to the wide variety of non-thermal populations (cosmic rays) that can contribute. Typically, the overall cosmic ray pressure in galaxy clusters is very low ($\lesssim 1\%$; \cite{Zandanel14,Bartels15,Pinzke17}) so ntSZ studies will be critical only within specific cluster regions, such as AGN bubbles for understanding the feedback mechanism \cite{Abdulla18,Lacy19}, or near cluster cores for finding the signature of annihilating dark-matter particles \cite{Colafrancesco04}.

\vspace{-3mm}
\vspace*{-9mm}

At the same time, sensitivities of the next generation CMB survey experiments can enable stacking analyses of the global ntSZ effect for an important class of objects: the radio halo clusters. These host Mpc-scale diffuse synchrotron emissions that are thought to be the signature of GeV-energy electrons energized by major mergers \cite{Cassano10,Brunetti14,Sommer14,vanWeeren19}, however this correspondence is highly uncertain and the determination of merger energetics via synchrotron emission is complicated by the unknown magnetic field strengths and topology. In the coming decade, all-sky radio surveys are expected to increase the number of radio halo clusters from $\sim50$ currently known objects to several hundreds \cite{Yuan15,Norris15,Knowles17}. 
A determination of their volume-averaged ntSZ signature will provide direct constraints on the energy dissipation and particle acceleration processes following major mergers, 
complementing the picture of a dynamical Universe as will be established by other probes.

\vspace{-2mm}
\section{The path forward in the coming decade}
\vspace{-2mm}
\textit{What are the instrumental requirements for the next generation to make significant progress?} \\
\vspace{-3mm}


\noindent 
The mm/submm community will move toward realizing the promises of SZ spectral science from both space and the ground.
New generation CMB and spectral intensity mapping surveys will carry on the rapid developments in microwave detector technology that are occurring right now, 
e.g., kinetic inductance detector (KID) based cameras coupled with imaging interferometers like CONCERTO \cite{Lagache18} and more sensitive future cameras with significantly higher optical throughput like Prime-Cam \cite{Vavagiakis18} (the latter hosting multichroic detectors with an imaging Fabry-P\'erot interferometer \cite{Stacey2018}). 
Efficient instantaneous sampling of the spectrum can be possible with 
broadband multichroic focal planes consisting of multimoded feeds \cite{Johnson2018}, 
multi-scale antenna arrays \cite{Ji2014}, 
and direct detection spectrometers like TIME \cite{TIME-Pilot}, Wspec \cite{Bryan2016}, Micro-Spec \cite{Barrentine2016} 
and DESHIMA \cite{Endo2012}. 

CCAT-prime, located at 5600~m altitude in the Atacama, will be among the first survey telescopes in the coming decade to scan a large fraction of the sky in the submm for CMB science \cite{Parshley18}. It will also provide ample scope for future generation instruments to take advantage of this excellent site. Other locations with comparable atmospheric transmission are at the South Pole or Dome-C in Antarctica, where for example the current 10-m SPT dish is already used for submm VLBI \cite{Kim18SPT}. Future developments in the Atacama desert will likely include the CSST \cite{Padin14} and lead up to  AtLAST  \cite{Bertoldi18,deBreuck18,Mroczkowski18}, a 50-m class submm telescope that will revolutionize SZ spectral science. 
Looking from space, future CMB missions like PICO \cite{pico18} and CMB-Bharat \cite{CMB-Bharat} --
building upon the scientifically compelling mission concepts of PRISM \cite{PRISM2013} and CORE \cite{COREmission} --  
will provide full-sky coverage with angular resolution similar to {\sl Planck}'s, but with more spectral channels and much better sensitivity. Those will be complemented by infra-red missions like SPICA \cite{Spica16}, to deliver THz-frequency data with similarly high-precision for  modeling the thermal dust emission. 



\bibliographystyle{unsrturltrunc6}
\bibliography{szrefs}

\begin{thebibliography}{10}

\bibitem{Allen11}
S.~W. {Allen}, A.~E. {Evrard}, and A.~B. {Mantz}.
\newblock {Cosmological Parameters from Observations of Galaxy Clusters}.
\newblock {\em \araa}, 49:409--470, September 2011.
\newblock \href {http://arxiv.org/abs/1103.4829} {\path{arXiv:1103.4829}},
  \href {http://dx.doi.org/10.1146/annurev-astro-081710-102514}
  {\path{doi:10.1146/annurev-astro-081710-102514}}.

\bibitem{Kravtsov12}
A.~V. {Kravtsov} and S.~{Borgani}.
\newblock {Formation of Galaxy Clusters}.
\newblock {\em \araa}, 50:353--409, September 2012.
\newblock \href {http://arxiv.org/abs/1205.5556} {\path{arXiv:1205.5556}},
  \href {http://dx.doi.org/10.1146/annurev-astro-081811-125502}
  {\path{doi:10.1146/annurev-astro-081811-125502}}.

\bibitem{Nagai14}
D.~{Nagai}.
\newblock {Cosmology and astrophysics with galaxy clusters}.
\newblock In {\em American Institute of Physics Conference Series}, volume 1632
  of {\em American Institute of Physics Conference Series}, pages 88--106,
  November 2014.
\newblock \href {http://dx.doi.org/10.1063/1.4902845}
  {\path{doi:10.1063/1.4902845}}.

\bibitem{Sarazin86}
C.~L. {Sarazin}.
\newblock {X-ray emission from clusters of galaxies}.
\newblock {\em Reviews of Modern Physics}, 58:1--115, January 1986.
\newblock \href {http://dx.doi.org/10.1103/RevModPhys.58.1}
  {\path{doi:10.1103/RevModPhys.58.1}}.

\bibitem{Kormendy89}
J.~{Kormendy} and S.~{Djorgovski}.
\newblock {Surface photometry and the structure of elliptical galaxies}.
\newblock {\em \araa}, 27:235--277, 1989.
\newblock \href {http://dx.doi.org/10.1146/annurev.aa.27.090189.001315}
  {\path{doi:10.1146/annurev.aa.27.090189.001315}}.

\bibitem{Bykov00}
A.~M. {Bykov}, H.~{Bloemen}, and Y.~A. {Uvarov}.
\newblock {Nonthermal emission from clusters of galaxies}.
\newblock {\em \aap}, 362:886--894, October 2000.

\bibitem{SZ70}
R.~A. {Sunyaev} and Y.~B. {Zeldovich}.
\newblock {The Spectrum of Primordial Radiation, its Distortions and their
  Significance}.
\newblock {\em Comments on Astrophysics and Space Physics}, 2:66, March 1970.

\bibitem{SZ72}
R.~A. {Sunyaev} and Y.~B. {Zeldovich}.
\newblock {The Observations of Relic Radiation as a Test of the Nature of X-Ray
  Radiation from the Clusters of Galaxies}.
\newblock {\em Comments on Astrophysics and Space Physics}, 4:173, November
  1972.

\bibitem{SZ75}
R.~A. {Sunyaev} and Y.~B. {Zeldovich}.
\newblock {The velocity of clusters of galaxies relative to the microwave
  background. The possibility of its measurement}.
\newblock {\em NASA STI/Recon Technical Report N}, 80, 1975.

\bibitem{SZ80rev}
R.~A. {Sunyaev} and I.~B. {Zeldovich}.
\newblock {Microwave background radiation as a probe of the contemporary
  structure and history of the universe}.
\newblock {\em \araa}, 18:537--560, 1980.
\newblock \href {http://dx.doi.org/10.1146/annurev.aa.18.090180.002541}
  {\path{doi:10.1146/annurev.aa.18.090180.002541}}.

\bibitem{Birkinshaw99}
M.~{Birkinshaw}.
\newblock {The Sunyaev-Zel'dovich effect}.
\newblock {\em \physrep}, 310:97--195, March 1999.
\newblock \href {http://arxiv.org/abs/astro-ph/9808050}
  {\path{arXiv:astro-ph/9808050}}, \href
  {http://dx.doi.org/10.1016/S0370-1573(98)00080-5}
  {\path{doi:10.1016/S0370-1573(98)00080-5}}.

\bibitem{Carlstrom02}
J.~E. {Carlstrom}, G.~P. {Holder}, and E.~D. {Reese}.
\newblock {Cosmology with the Sunyaev-Zel'dovich Effect}.
\newblock {\em \araa}, 40:643--680, 2002.
\newblock \href {http://arxiv.org/abs/astro-ph/0208192}
  {\path{arXiv:astro-ph/0208192}}, \href
  {http://dx.doi.org/10.1146/annurev.astro.40.060401.093803}
  {\path{doi:10.1146/annurev.astro.40.060401.093803}}.

\bibitem{SSRev2019}
T.~{Mroczkowski}, D.~{Nagai}, K.~{Basu}, J.~{Chluba}, J.~{Sayers}, R.~{Adam},
  et~al.
\newblock {Astrophysics with the Spatially and Spectrally Resolved
  Sunyaev-Zeldovich Effects. A Millimetre/Submillimetre Probe of the Warm and
  Hot Universe}.
\newblock {\em \ssr}, 215:17, February 2019.
\newblock \href {http://arxiv.org/abs/1811.02310} {\path{arXiv:1811.02310}},
  \href {http://dx.doi.org/10.1007/s11214-019-0581-2}
  {\path{doi:10.1007/s11214-019-0581-2}}.

\bibitem{Stanis09}
Z.~{Staniszewski}, P.~A.~R. {Ade}, K.~A. {Aird}, B.~A. {Benson}, L.~E. {Bleem},
  J.~E. {Carlstrom}, et~al.
\newblock {Galaxy Clusters Discovered with a Sunyaev-Zel'dovich Effect Survey}.
\newblock {\em \apj}, 701:32--41, August 2009.
\newblock \href {http://arxiv.org/abs/0810.1578} {\path{arXiv:0810.1578}},
  \href {http://dx.doi.org/10.1088/0004-637X/701/1/32}
  {\path{doi:10.1088/0004-637X/701/1/32}}.

\bibitem{Menanteau10}
F.~{Menanteau}, J.~{Gonz{\'a}lez}, J.-B. {Juin}, T.~A. {Marriage}, E.~D.
  {Reese}, V.~{Acquaviva}, et~al.
\newblock {The Atacama Cosmology Telescope: Physical Properties and Purity of a
  Galaxy Cluster Sample Selected via the Sunyaev-Zel'dovich Effect}.
\newblock {\em \apj}, 723:1523--1541, November 2010.
\newblock \href {http://arxiv.org/abs/1006.5126} {\path{arXiv:1006.5126}},
  \href {http://dx.doi.org/10.1088/0004-637X/723/2/1523}
  {\path{doi:10.1088/0004-637X/723/2/1523}}.

\bibitem{SOscience}
{The Simons Observatory Collaboration}, P.~{Ade}, J.~{Aguirre}, Z.~{Ahmed},
  S.~{Aiola}, A.~{Ali}, et~al.
\newblock {The Simons Observatory: Science goals and forecasts}.
\newblock {\em arXiv e-prints}, August 2018.
\newblock \href {http://arxiv.org/abs/1808.07445} {\path{arXiv:1808.07445}}.

\bibitem{CMB-S4}
K.~N. {Abazajian}, P.~{Adshead}, Z.~{Ahmed}, S.~W. {Allen}, D.~{Alonso}, K.~S.
  {Arnold}, et~al.
\newblock {CMB-S4 Science Book, First Edition}.
\newblock {\em arXiv e-prints}, October 2016.
\newblock \href {http://arxiv.org/abs/1610.02743} {\path{arXiv:1610.02743}}.

\bibitem{COREclust}
J.-B. {Melin}, A.~{Bonaldi}, M.~{Remazeilles}, S.~{Hagstotz}, J.~M. {Diego},
  C.~{Hern{\'a}ndez-Monteagudo}, et~al.
\newblock {Exploring cosmic origins with CORE: Cluster science}.
\newblock {\em \jcap}, 4:019, April 2018.
\newblock \href {http://arxiv.org/abs/1703.10456} {\path{arXiv:1703.10456}},
  \href {http://dx.doi.org/10.1088/1475-7516/2018/04/019}
  {\path{doi:10.1088/1475-7516/2018/04/019}}.

\bibitem{Lamarre98}
J.~M. {Lamarre}, M.~{Giard}, E.~{Pointecouteau}, J.~P. {Bernard}, G.~{Serra},
  F.~{Pajot}, et~al.
\newblock {First Measurement of the Submillimeter Sunyaev-Zeldovich Effect}.
\newblock {\em \apjl}, 507:L5--L8, November 1998.
\newblock \href {http://arxiv.org/abs/astro-ph/9806128}
  {\path{arXiv:astro-ph/9806128}}, \href {http://dx.doi.org/10.1086/311678}
  {\path{doi:10.1086/311678}}.

\bibitem{Komatsu99}
E.~{Komatsu}, T.~{Kitayama}, Y.~{Suto}, M.~{Hattori}, R.~{Kawabe}, H.~{Matsuo},
  et~al.
\newblock {Submillimeter Detection of the Sunyaev-Zeldovich Effect toward the
  Most Luminous X-Ray Cluster at Z = 0.45}.
\newblock {\em \apjl}, 516:L1--L4, May 1999.
\newblock \href {http://arxiv.org/abs/astro-ph/9902351}
  {\path{arXiv:astro-ph/9902351}}, \href {http://dx.doi.org/10.1086/311983}
  {\path{doi:10.1086/311983}}.

\bibitem{DeDeo05}
S.~{DeDeo}, D.~N. {Spergel}, and H.~{Trac}.
\newblock {The kinetic Sunyaev-Zel'dovitch effect as a dark energy probe}.
\newblock {\em arXiv Astrophysics e-prints}, November 2005.
\newblock \href {http://arxiv.org/abs/astro-ph/0511060}
  {\path{arXiv:astro-ph/0511060}}.

\bibitem{Bhatta08}
S.~{Bhattacharya} and A.~{Kosowsky}.
\newblock {Dark energy constraints from galaxy cluster peculiar velocities}.
\newblock {\em \prd}, 77(8):083004, April 2008.
\newblock \href {http://arxiv.org/abs/0712.0034} {\path{arXiv:0712.0034}},
  \href {http://dx.doi.org/10.1103/PhysRevD.77.083004}
  {\path{doi:10.1103/PhysRevD.77.083004}}.

\bibitem{Mueller15}
E.-M. {Mueller}, F.~{de Bernardis}, R.~{Bean}, and M.~D. {Niemack}.
\newblock {Constraints on Gravity and Dark Energy from the Pairwise Kinematic
  Sunyaev-Zel'dovich Effect}.
\newblock {\em \apj}, 808:47, July 2015.
\newblock \href {http://arxiv.org/abs/1408.6248} {\path{arXiv:1408.6248}},
  \href {http://dx.doi.org/10.1088/0004-637X/808/1/47}
  {\path{doi:10.1088/0004-637X/808/1/47}}.

\bibitem{Schaan16}
E.~{Schaan}, S.~{Ferraro}, M.~{Vargas-Maga{\~n}a}, K.~M. {Smith}, S.~{Ho},
  S.~{Aiola}, et~al.
\newblock {Evidence for the kinematic Sunyaev-Zel'dovich effect with the
  Atacama Cosmology Telescope and velocity reconstruction from the Baryon
  Oscillation Spectroscopic Survey}.
\newblock {\em \prd}, 93(8):082002, April 2016.
\newblock \href {http://arxiv.org/abs/1510.06442} {\path{arXiv:1510.06442}},
  \href {http://dx.doi.org/10.1103/PhysRevD.93.082002}
  {\path{doi:10.1103/PhysRevD.93.082002}}.

\bibitem{Sugiyama17}
N.~S. {Sugiyama}, T.~{Okumura}, and D.~N. {Spergel}.
\newblock {Will kinematic Sunyaev-Zel'dovich measurements enhance the science
  return from galaxy redshift surveys?}
\newblock {\em \jcap}, 1:057, January 2017.
\newblock \href {http://arxiv.org/abs/1606.06367} {\path{arXiv:1606.06367}},
  \href {http://dx.doi.org/10.1088/1475-7516/2017/01/057}
  {\path{doi:10.1088/1475-7516/2017/01/057}}.

\bibitem{Mroczkowski12}
T.~{Mroczkowski}, S.~{Dicker}, J.~{Sayers}, E.~D. {Reese}, B.~{Mason},
  N.~{Czakon}, et~al.
\newblock {A Multi-wavelength Study of the Sunyaev-Zel'dovich Effect in the
  Triple-merger Cluster MACS J0717.5+3745 with MUSTANG and Bolocam}.
\newblock {\em \apj}, 761:47, December 2012.
\newblock \href {http://arxiv.org/abs/1205.0052} {\path{arXiv:1205.0052}},
  \href {http://dx.doi.org/10.1088/0004-637X/761/1/47}
  {\path{doi:10.1088/0004-637X/761/1/47}}.

\bibitem{Sayers13}
J.~{Sayers}, T.~{Mroczkowski}, M.~{Zemcov}, P.~M. {Korngut}, J.~{Bock},
  E.~{Bulbul}, et~al.
\newblock {A Measurement of the Kinetic Sunyaev-Zel'dovich Signal Toward MACS
  J0717.5+3745}.
\newblock {\em \apj}, 778:52, November 2013.
\newblock \href {http://arxiv.org/abs/1312.3680} {\path{arXiv:1312.3680}},
  \href {http://dx.doi.org/10.1088/0004-637X/778/1/52}
  {\path{doi:10.1088/0004-637X/778/1/52}}.

\bibitem{Adam17}
R.~{Adam}, M.~{Arnaud}, I.~{Bartalucci}, P.~{Ade}, P.~{Andr{\'e}}, A.~{Beelen},
  et~al.
\newblock {Mapping the hot gas temperature in galaxy clusters using X-ray and
  Sunyaev-Zel'dovich imaging}.
\newblock {\em \aap}, 606:A64, October 2017.
\newblock \href {http://arxiv.org/abs/1706.10230} {\path{arXiv:1706.10230}},
  \href {http://dx.doi.org/10.1051/0004-6361/201629810}
  {\path{doi:10.1051/0004-6361/201629810}}.

\bibitem{Hand12}
N.~{Hand}, G.~E. {Addison}, E.~{Aubourg}, N.~{Battaglia}, E.~S. {Battistelli},
  D.~{Bizyaev}, et~al.
\newblock {Evidence of Galaxy Cluster Motions with the Kinematic
  Sunyaev-Zel'dovich Effect}.
\newblock {\em Physical Review Letters}, 109(4):041101, July 2012.
\newblock \href {http://arxiv.org/abs/1203.4219} {\path{arXiv:1203.4219}},
  \href {http://dx.doi.org/10.1103/PhysRevLett.109.041101}
  {\path{doi:10.1103/PhysRevLett.109.041101}}.

\bibitem{PlXIII14}
{Planck Collaboration}, P.~A.~R. {Ade}, N.~{Aghanim}, M.~{Arnaud},
  M.~{Ashdown}, J.~{Aumont}, et~al.
\newblock {Planck intermediate results. XIII. Constraints on peculiar
  velocities}.
\newblock {\em \aap}, 561:A97, January 2014.
\newblock \href {http://arxiv.org/abs/1303.5090} {\path{arXiv:1303.5090}},
  \href {http://dx.doi.org/10.1051/0004-6361/201321299}
  {\path{doi:10.1051/0004-6361/201321299}}.

\bibitem{Soergel18}
B.~{Soergel}, A.~{Saro}, T.~{Giannantonio}, G.~{Efstathiou}, and K.~{Dolag}.
\newblock {Cosmology with the pairwise kinematic SZ effect: calibration and
  validation using hydrodynamical simulations}.
\newblock {\em \mnras}, 478:5320--5335, August 2018.
\newblock \href {http://arxiv.org/abs/1712.05714} {\path{arXiv:1712.05714}},
  \href {http://dx.doi.org/10.1093/mnras/sty1324}
  {\path{doi:10.1093/mnras/sty1324}}.

\bibitem{PlXXXVII16}
{Planck Collaboration}, P.~A.~R. {Ade}, N.~{Aghanim}, M.~{Arnaud},
  M.~{Ashdown}, E.~{Aubourg}, et~al.
\newblock {Planck intermediate results. XXXVII. Evidence of unbound gas from
  the kinetic Sunyaev-Zeldovich effect}.
\newblock {\em \aap}, 586:A140, February 2016.
\newblock \href {http://arxiv.org/abs/1504.03339} {\path{arXiv:1504.03339}},
  \href {http://dx.doi.org/10.1051/0004-6361/201526328}
  {\path{doi:10.1051/0004-6361/201526328}}.

\bibitem{Hill16}
J.~C. {Hill}, S.~{Ferraro}, N.~{Battaglia}, J.~{Liu}, and D.~N. {Spergel}.
\newblock {Kinematic Sunyaev-Zel'dovich Effect with Projected Fields: A Novel
  Probe of the Baryon Distribution with Planck, WMAP, and WISE Data}.
\newblock {\em Physical Review Letters}, 117(5):051301, July 2016.
\newblock \href {http://arxiv.org/abs/1603.01608} {\path{arXiv:1603.01608}},
  \href {http://dx.doi.org/10.1103/PhysRevLett.117.051301}
  {\path{doi:10.1103/PhysRevLett.117.051301}}.

\bibitem{Lim17}
S.~{Lim}, H.~{Mo}, H.~{Wang}, and X.~{Yang}.
\newblock {The detection of missing baryons in galaxy halos with kinetic
  Sunyaev-Zel'dovich effect}.
\newblock {\em arXiv e-prints}, December 2017.
\newblock \href {http://arxiv.org/abs/1712.08619} {\path{arXiv:1712.08619}}.

\bibitem{Hurier17}
G.~{Hurier} and C.~{Tchernin}.
\newblock {Mapping the temperature of the intra-cluster medium with the thermal
  Sunyaev-Zel'dovich effect}.
\newblock {\em \aap}, 604:A94, August 2017.
\newblock \href {http://arxiv.org/abs/1702.03711} {\path{arXiv:1702.03711}},
  \href {http://dx.doi.org/10.1051/0004-6361/201629993}
  {\path{doi:10.1051/0004-6361/201629993}}.

\bibitem{Erler2018}
J.~{Erler}, K.~{Basu}, J.~{Chluba}, and F.~{Bertoldi}.
\newblock {Planck's view on the spectrum of the Sunyaev-Zeldovich effect}.
\newblock {\em \mnras}, 476:3360--3381, May 2018.
\newblock \href {http://arxiv.org/abs/1709.01187} {\path{arXiv:1709.01187}},
  \href {http://dx.doi.org/10.1093/mnras/sty327}
  {\path{doi:10.1093/mnras/sty327}}.

\bibitem{Mittal2018}
A.~{Mittal}, F.~{de Bernardis}, and M.~D. {Niemack}.
\newblock {Optimizing measurements of cluster velocities and temperatures for
  CCAT-prime and future surveys}.
\newblock {\em \jcap}, 2:032, February 2018.
\newblock \href {http://arxiv.org/abs/1708.06365} {\path{arXiv:1708.06365}},
  \href {http://dx.doi.org/10.1088/1475-7516/2018/02/032}
  {\path{doi:10.1088/1475-7516/2018/02/032}}.

\bibitem{Wright79}
E.~L. {Wright}.
\newblock {Distortion of the microwave background by a hot intergalactic
  medium}.
\newblock {\em \apj}, 232:348--351, September 1979.
\newblock \href {http://dx.doi.org/10.1086/157294} {\path{doi:10.1086/157294}}.

\bibitem{Pointecouteau98}
E.~{Pointecouteau}, M.~{Giard}, and D.~{Barret}.
\newblock {Determination of the hot intracluster gas temperature from
  submillimeter measurements}.
\newblock {\em \aap}, 336:44--48, August 1998.
\newblock \href {http://arxiv.org/abs/astro-ph/9712271}
  {\path{arXiv:astro-ph/9712271}}.

\bibitem{Nozawa09}
S.~{Nozawa}, Y.~{Kohyama}, and N.~{Itoh}.
\newblock {Study on the solutions of the Sunyaev-Zeldovich effect for clusters
  of galaxies}.
\newblock {\em \prd}, 79(12):123007, June 2009.
\newblock \href {http://arxiv.org/abs/0904.3811} {\path{arXiv:0904.3811}},
  \href {http://dx.doi.org/10.1103/PhysRevD.79.123007}
  {\path{doi:10.1103/PhysRevD.79.123007}}.

\bibitem{Chluba12}
J.~{Chluba}, D.~{Nagai}, S.~{Sazonov}, and K.~{Nelson}.
\newblock {A fast and accurate method for computing the Sunyaev-Zel'dovich
  signal of hot galaxy clusters}.
\newblock {\em \mnras}, 426:510--530, October 2012.
\newblock \href {http://arxiv.org/abs/1205.5778} {\path{arXiv:1205.5778}},
  \href {http://dx.doi.org/10.1111/j.1365-2966.2012.21741.x}
  {\path{doi:10.1111/j.1365-2966.2012.21741.x}}.

\bibitem{Hurier16}
G.~{Hurier}.
\newblock {High significance detection of the tSZ effect relativistic
  corrections}.
\newblock {\em \aap}, 596:A61, December 2016.
\newblock \href {http://arxiv.org/abs/1701.09020} {\path{arXiv:1701.09020}},
  \href {http://dx.doi.org/10.1051/0004-6361/201629726}
  {\path{doi:10.1051/0004-6361/201629726}}.

\bibitem{Borm14}
K.~{Borm}, T.~H. {Reiprich}, I.~{Mohammed}, and L.~{Lovisari}.
\newblock {Constraining galaxy cluster temperatures and redshifts with eROSITA
  survey data}.
\newblock {\em \aap}, 567:A65, July 2014.
\newblock \href {http://arxiv.org/abs/1404.5312} {\path{arXiv:1404.5312}},
  \href {http://dx.doi.org/10.1051/0004-6361/201322643}
  {\path{doi:10.1051/0004-6361/201322643}}.

\bibitem{Hofmann17}
F.~{Hofmann}, J.~S. {Sanders}, N.~{Clerc}, K.~{Nandra}, J.~{Ridl},
  K.~{Dennerl}, et~al.
\newblock {eROSITA cluster cosmology forecasts: Cluster temperature
  substructure bias}.
\newblock {\em \aap}, 606:A118, October 2017.
\newblock \href {http://arxiv.org/abs/1708.05205} {\path{arXiv:1708.05205}},
  \href {http://dx.doi.org/10.1051/0004-6361/201730742}
  {\path{doi:10.1051/0004-6361/201730742}}.

\bibitem{Hansen04}
S.~H. {Hansen}.
\newblock {Cluster temperature profiles and Sunyaev-Zeldovich observations}.
\newblock {\em \mnras}, 351:L5--L8, June 2004.
\newblock \href {http://arxiv.org/abs/astro-ph/0401391}
  {\path{arXiv:astro-ph/0401391}}, \href
  {http://dx.doi.org/10.1111/j.1365-2966.2004.07920.x}
  {\path{doi:10.1111/j.1365-2966.2004.07920.x}}.

\bibitem{Kay08}
S.~T. {Kay}, L.~C. {Powell}, A.~R. {Liddle}, and P.~A. {Thomas}.
\newblock {The Sunyaev-Zel'dovich temperature of the intracluster medium}.
\newblock {\em \mnras}, 386:2110--2114, June 2008.
\newblock \href {http://arxiv.org/abs/0706.3668} {\path{arXiv:0706.3668}},
  \href {http://dx.doi.org/10.1111/j.1365-2966.2008.13183.x}
  {\path{doi:10.1111/j.1365-2966.2008.13183.x}}.

\bibitem{Mazzotta04}
P.~{Mazzotta}, E.~{Rasia}, L.~{Moscardini}, and G.~{Tormen}.
\newblock {Comparing the temperatures of galaxy clusters from hydrodynamical
  N-body simulations to Chandra and XMM-Newton observations}.
\newblock {\em \mnras}, 354:10--24, October 2004.
\newblock \href {http://arxiv.org/abs/astro-ph/0404425}
  {\path{arXiv:astro-ph/0404425}}, \href
  {http://dx.doi.org/10.1111/j.1365-2966.2004.08167.x}
  {\path{doi:10.1111/j.1365-2966.2004.08167.x}}.

\bibitem{Remazeilles2019}
M.~{Remazeilles}, B.~{Bolliet}, A.~{Rotti}, and J.~{Chluba}.
\newblock {Can we neglect relativistic temperature corrections in the Planck
  thermal SZ analysis?}
\newblock {\em \mnras}, 483:3459--3464, March 2019.
\newblock \href {http://arxiv.org/abs/1809.09666} {\path{arXiv:1809.09666}},
  \href {http://dx.doi.org/10.1093/mnras/sty3352}
  {\path{doi:10.1093/mnras/sty3352}}.

\bibitem{PlDust1}
{Planck Collaboration}, P.~A.~R. {Ade}, N.~{Aghanim}, M.~{Arnaud}, J.~{Aumont},
  C.~{Baccigalupi}, et~al.
\newblock {Planck 2015 results. XXIII. The thermal Sunyaev-Zeldovich
  effect-cosmic infrared background correlation}.
\newblock {\em \aap}, 594:A23, September 2016.
\newblock \href {http://arxiv.org/abs/1509.06555} {\path{arXiv:1509.06555}},
  \href {http://dx.doi.org/10.1051/0004-6361/201527418}
  {\path{doi:10.1051/0004-6361/201527418}}.

\bibitem{PlDust2}
{Planck Collaboration}, R.~{Adam}, P.~A.~R. {Ade}, N.~{Aghanim}, M.~{Ashdown},
  J.~{Aumont}, et~al.
\newblock {Planck intermediate results. XLIII. Spectral energy distribution of
  dust in clusters of galaxies}.
\newblock {\em \aap}, 596:A104, December 2016.
\newblock \href {http://arxiv.org/abs/1603.04919} {\path{arXiv:1603.04919}},
  \href {http://dx.doi.org/10.1051/0004-6361/201628522}
  {\path{doi:10.1051/0004-6361/201628522}}.

\bibitem{Melin2018}
J.-B. {Melin}, J.~G. {Bartlett}, Z.-Y. {Cai}, G.~{De Zotti}, J.~{Delabrouille},
  M.~{Roman}, et~al.
\newblock {Dust in galaxy clusters: Modeling at millimeter wavelengths and
  impact on Planck cluster cosmology}.
\newblock {\em \aap}, 617:A75, September 2018.
\newblock \href {http://arxiv.org/abs/1808.06807} {\path{arXiv:1808.06807}},
  \href {http://dx.doi.org/10.1051/0004-6361/201732292}
  {\path{doi:10.1051/0004-6361/201732292}}.

\bibitem{Vogelsberger18}
M.~{Vogelsberger}, R.~{McKinnon}, S.~{O'Neil}, F.~{Marinacci}, P.~{Torrey}, and
  R.~{Kannan}.
\newblock {Dust in and around galaxies: dust in cluster environments and its
  impact on gas cooling}.
\newblock {\em arXiv e-prints}, November 2018.
\newblock \href {http://arxiv.org/abs/1811.05477} {\path{arXiv:1811.05477}}.

\bibitem{Werner2010}
N.~{Werner}, A.~{Simionescu}, E.~T. {Million}, S.~W. {Allen}, P.~E.~J.
  {Nulsen}, A.~{von der Linden}, et~al.
\newblock {Feedback under the microscope-II. Heating, gas uplift and mixing in
  the nearest cluster core}.
\newblock {\em \mnras}, 407:2063--2074, October 2010.
\newblock \href {http://arxiv.org/abs/1003.5334} {\path{arXiv:1003.5334}},
  \href {http://dx.doi.org/10.1111/j.1365-2966.2010.16755.x}
  {\path{doi:10.1111/j.1365-2966.2010.16755.x}}.

\bibitem{Soergel2017}
B.~{Soergel}, T.~{Giannantonio}, G.~{Efstathiou}, E.~{Puchwein}, and
  D.~{Sijacki}.
\newblock {Constraints on AGN feedback from its Sunyaev-Zel'dovich imprint on
  the cosmic background radiation}.
\newblock {\em \mnras}, 468:577--596, June 2017.
\newblock \href {http://arxiv.org/abs/1612.06296} {\path{arXiv:1612.06296}},
  \href {http://dx.doi.org/10.1093/mnras/stx492}
  {\path{doi:10.1093/mnras/stx492}}.

\bibitem{Battaglia16}
N.~{Battaglia}.
\newblock {The tau of galaxy clusters}.
\newblock {\em \jcap}, 8:058, August 2016.
\newblock \href {http://arxiv.org/abs/1607.02442} {\path{arXiv:1607.02442}},
  \href {http://dx.doi.org/10.1088/1475-7516/2016/08/058}
  {\path{doi:10.1088/1475-7516/2016/08/058}}.

\bibitem{Flender17}
S.~{Flender}, D.~{Nagai}, and M.~{McDonald}.
\newblock {Constraints on the Optical Depth of Galaxy Groups and Clusters}.
\newblock {\em \apj}, 837:124, March 2017.
\newblock \href {http://arxiv.org/abs/1610.08029} {\path{arXiv:1610.08029}},
  \href {http://dx.doi.org/10.3847/1538-4357/aa60bf}
  {\path{doi:10.3847/1538-4357/aa60bf}}.

\bibitem{Stacey2018}
G.~J. {Stacey}, M.~{Aravena}, K.~{Basu}, N.~{Battaglia}, B.~{Beringue},
  F.~{Bertoldi}, et~al.
\newblock {CCAT-Prime: science with an ultra-widefield submillimeter
  observatory on Cerro Chajnantor}.
\newblock In {\em Ground-based and Airborne Telescopes VII}, volume 10700 of
  {\em Society of Photo-Optical Instrumentation Engineers (SPIE) Conference
  Series}, page 107001M, July 2018.
\newblock \href {http://arxiv.org/abs/1807.04354} {\path{arXiv:1807.04354}},
  \href {http://dx.doi.org/10.1117/12.2314031} {\path{doi:10.1117/12.2314031}}.

\bibitem{Chen2018}
T.~{Chen}, M.~{Remazeilles}, and C.~{Dickinson}.
\newblock {Impact of SZ cluster residuals in CMB maps and CMB-LSS
  cross-correlations}.
\newblock {\em \mnras}, 479:4239--4252, September 2018.
\newblock \href {http://arxiv.org/abs/1803.08853} {\path{arXiv:1803.08853}},
  \href {http://dx.doi.org/10.1093/mnras/sty1730}
  {\path{doi:10.1093/mnras/sty1730}}.

\bibitem{Alonso16}
D.~{Alonso}, T.~{Louis}, P.~{Bull}, and P.~G. {Ferreira}.
\newblock {Reconstructing cosmic growth with kinetic Sunyaev-Zel'dovich
  observations in the era of stage IV experiments}.
\newblock {\em \prd}, 94(4):043522, August 2016.
\newblock \href {http://arxiv.org/abs/1604.01382} {\path{arXiv:1604.01382}},
  \href {http://dx.doi.org/10.1103/PhysRevD.94.043522}
  {\path{doi:10.1103/PhysRevD.94.043522}}.

\bibitem{Hernandez08}
C.~{Hern{\'a}ndez-Monteagudo} and R.~A. {Sunyaev}.
\newblock {Missing baryons, bulk flows, and the E-mode polarization of the
  Cosmic Microwave Background}.
\newblock {\em \aap}, 490:25--29, October 2008.
\newblock \href {http://arxiv.org/abs/0805.3702} {\path{arXiv:0805.3702}},
  \href {http://dx.doi.org/10.1051/0004-6361:200810204}
  {\path{doi:10.1051/0004-6361:200810204}}.

\bibitem{Battaglia17}
N.~{Battaglia}, S.~{Ferraro}, E.~{Schaan}, and D.~N. {Spergel}.
\newblock {Future constraints on halo thermodynamics from combined
  Sunyaev-Zel'dovich measurements}.
\newblock {\em \jcap}, 11:040, November 2017.
\newblock \href {http://arxiv.org/abs/1705.05881} {\path{arXiv:1705.05881}},
  \href {http://dx.doi.org/10.1088/1475-7516/2017/11/040}
  {\path{doi:10.1088/1475-7516/2017/11/040}}.

\bibitem{Rephaeli95}
Y.~{Rephaeli}.
\newblock {Cosmic microwave background comptonization by hot intracluster gas}.
\newblock {\em \apj}, 445:33--36, May 1995.
\newblock \href {http://dx.doi.org/10.1086/175669} {\path{doi:10.1086/175669}}.

\bibitem{Enss00}
T.~A. {En{\ss}lin} and C.~R. {Kaiser}.
\newblock {Comptonization of the cosmic microwave background by relativistic
  plasma}.
\newblock {\em \aap}, 360:417--430, August 2000.
\newblock \href {http://arxiv.org/abs/arXiv:astro-ph/0001429}
  {\path{arXiv:arXiv:astro-ph/0001429}}.

\bibitem{Colafrancesco03}
S.~{Colafrancesco}, P.~{Marchegiani}, and E.~{Palladino}.
\newblock {The non-thermal Sunyaev-Zel'dovich effect in clusters of galaxies}.
\newblock {\em \aap}, 397:27--52, January 2003.
\newblock \href {http://arxiv.org/abs/arXiv:astro-ph/0211649}
  {\path{arXiv:arXiv:astro-ph/0211649}}, \href
  {http://dx.doi.org/10.1051/0004-6361:20021199}
  {\path{doi:10.1051/0004-6361:20021199}}.

\bibitem{Zandanel14}
F.~{Zandanel}, C.~{Pfrommer}, and F.~{Prada}.
\newblock {On the physics of radio haloes in galaxy clusters: scaling relations
  and luminosity functions}.
\newblock {\em \mnras}, 438:124--144, February 2014.
\newblock \href {http://arxiv.org/abs/1311.4795} {\path{arXiv:1311.4795}},
  \href {http://dx.doi.org/10.1093/mnras/stt2250}
  {\path{doi:10.1093/mnras/stt2250}}.

\bibitem{Bartels15}
R.~{Bartels}, F.~{Zandanel}, and S.~{Ando}.
\newblock {Inverse-Compton emission from clusters of galaxies: Predictions for
  ASTRO-H}.
\newblock {\em \aap}, 582:A20, September 2015.
\newblock \href {http://arxiv.org/abs/1501.06940} {\path{arXiv:1501.06940}},
  \href {http://dx.doi.org/10.1051/0004-6361/201525758}
  {\path{doi:10.1051/0004-6361/201525758}}.

\bibitem{Pinzke17}
A.~{Pinzke}, S.~P. {Oh}, and C.~{Pfrommer}.
\newblock {Turbulence and particle acceleration in giant radio haloes: the
  origin of seed electrons}.
\newblock {\em \mnras}, 465:4800--4816, March 2017.
\newblock \href {http://arxiv.org/abs/1611.07533} {\path{arXiv:1611.07533}},
  \href {http://dx.doi.org/10.1093/mnras/stw3024}
  {\path{doi:10.1093/mnras/stw3024}}.

\bibitem{Abdulla18}
Z.~{Abdulla}, J.~E. {Carlstrom}, A.~B. {Mantz}, D.~P. {Marrone}, C.~H. {Greer},
  J.~W. {Lamb}, et~al.
\newblock {Constraints on the Thermal Contents of the X-ray Cavities of Cluster
  MS 0735.6+7421 with Sunyaev-Zel'dovich Effect Observations}.
\newblock {\em ArXiv e-prints}, page arxiv:1806.05050, June 2018.
\newblock \href {http://arxiv.org/abs/1806.05050} {\path{arXiv:1806.05050}}.

\bibitem{Lacy19}
M.~{Lacy}, B.~{Mason}, C.~{Sarazin}, S.~{Chatterjee}, K.~{Nyland},
  A.~{Kimball}, et~al.
\newblock {Direct detection of quasar feedback via the Sunyaev-Zeldovich
  effect}.
\newblock {\em \mnras}, 483:L22--L27, February 2019.
\newblock \href {http://arxiv.org/abs/1811.05023} {\path{arXiv:1811.05023}},
  \href {http://dx.doi.org/10.1093/mnrasl/sly215}
  {\path{doi:10.1093/mnrasl/sly215}}.

\bibitem{Colafrancesco04}
S.~{Colafrancesco}.
\newblock {SZ effect from Dark Matter annihilation}.
\newblock {\em \aap}, 422:L23--L27, July 2004.
\newblock \href {http://arxiv.org/abs/astro-ph/0405456}
  {\path{arXiv:astro-ph/0405456}}, \href
  {http://dx.doi.org/10.1051/0004-6361:20040175}
  {\path{doi:10.1051/0004-6361:20040175}}.

\bibitem{Cassano10}
R.~{Cassano}, S.~{Ettori}, S.~{Giacintucci}, G.~{Brunetti}, M.~{Markevitch},
  T.~{Venturi}, et~al.
\newblock {On the Connection Between Giant Radio Halos and Cluster Mergers}.
\newblock {\em \apjl}, 721:L82--L85, October 2010.
\newblock \href {http://arxiv.org/abs/1008.3624} {\path{arXiv:1008.3624}},
  \href {http://dx.doi.org/10.1088/2041-8205/721/2/L82}
  {\path{doi:10.1088/2041-8205/721/2/L82}}.

\bibitem{Brunetti14}
G.~{Brunetti} and T.~W. {Jones}.
\newblock {Cosmic Rays in Galaxy Clusters and Their Nonthermal Emission}.
\newblock {\em International Journal of Modern Physics D}, 23:1430007--98,
  March 2014.
\newblock \href {http://arxiv.org/abs/1401.7519} {\path{arXiv:1401.7519}},
  \href {http://dx.doi.org/10.1142/S0218271814300079}
  {\path{doi:10.1142/S0218271814300079}}.

\bibitem{Sommer14}
M.~W. {Sommer} and K.~{Basu}.
\newblock {A comparative study of radio halo occurrence in SZ and X-ray
  selected galaxy cluster samples}.
\newblock {\em \mnras}, 437:2163--2179, January 2014.
\newblock \href {http://arxiv.org/abs/1307.3049} {\path{arXiv:1307.3049}},
  \href {http://dx.doi.org/10.1093/mnras/stt2037}
  {\path{doi:10.1093/mnras/stt2037}}.

\bibitem{vanWeeren19}
R.~J. {van Weeren}, F.~{de Gasperin}, H.~{Akamatsu}, M.~{Br{\"u}ggen},
  L.~{Feretti}, H.~{Kang}, et~al.
\newblock {Diffuse Radio Emission from Galaxy Clusters}.
\newblock {\em \ssr}, 215:16, February 2019.
\newblock \href {http://arxiv.org/abs/1901.04496} {\path{arXiv:1901.04496}},
  \href {http://dx.doi.org/10.1007/s11214-019-0584-z}
  {\path{doi:10.1007/s11214-019-0584-z}}.

\bibitem{Yuan15}
Z.~S. {Yuan}, J.~L. {Han}, and Z.~L. {Wen}.
\newblock {The Scaling Relations and the Fundamental Plane for Radio Halos and
  Relics of Galaxy Clusters}.
\newblock {\em \apj}, 813:77, November 2015.
\newblock \href {http://arxiv.org/abs/1510.04980} {\path{arXiv:1510.04980}},
  \href {http://dx.doi.org/10.1088/0004-637X/813/1/77}
  {\path{doi:10.1088/0004-637X/813/1/77}}.

\bibitem{Norris15}
R.~{Norris}, K.~{Basu}, M.~{Brown}, E.~{Carretti}, A.~D. {Kapinska},
  I.~{Prandoni}, et~al.
\newblock {The SKA Mid-frequency All-sky Continuum Survey: Discovering the
  unexpected and transforming radio-astronomy}.
\newblock {\em Advancing Astrophysics with the Square Kilometre Array
  (AASKA14)}, page~86, April 2015.
\newblock \href {http://arxiv.org/abs/1412.6076} {\path{arXiv:1412.6076}}.

\bibitem{Knowles17}
K.~{Knowles}, A.~{Baker}, K.~{Basu}, V.~{Bharadwaj}, R.~{Deane}, M.~{Devlin},
  et~al.
\newblock {MERGHERS: An SZ-selected cluster survey with MeerKAT}.
\newblock {\em arXiv e-prints}, September 2017.
\newblock \href {http://arxiv.org/abs/1709.03318} {\path{arXiv:1709.03318}}.

\bibitem{Lagache18}
G.~{Lagache}.
\newblock {Exploring the dusty star-formation in the early Universe using
  intensity mapping}.
\newblock In V.~{Jeli{\'c}} and T.~{van der Hulst}, editors, {\em IAU
  Symposium}, volume 333 of {\em IAU Symposium}, pages 228--233, May 2018.
\newblock \href {http://arxiv.org/abs/1801.08054} {\path{arXiv:1801.08054}},
  \href {http://dx.doi.org/10.1017/S1743921318000558}
  {\path{doi:10.1017/S1743921318000558}}.

\bibitem{Vavagiakis18}
E.~M. {Vavagiakis}, Z.~{Ahmed}, A.~{Ali}, K.~{Basu}, N.~{Battaglia},
  F.~{Bertoldi}, et~al.
\newblock {Prime-Cam: a first-light instrument for the CCAT-prime telescope}.
\newblock In {\em Millimeter, Submillimeter, and Far-Infrared Detectors and
  Instrumentation for Astronomy IX}, volume 10708 of {\em Society of
  Photo-Optical Instrumentation Engineers (SPIE) Conference Series}, page
  107081U, July 2018.
\newblock \href {http://arxiv.org/abs/1807.00058} {\path{arXiv:1807.00058}},
  \href {http://dx.doi.org/10.1117/12.2313868} {\path{doi:10.1117/12.2313868}}.

\bibitem{Johnson2018}
B.~R. {Johnson}, D.~{Flanigan}, M.~H. {Abitbol}, P.~A.~R. {Ade}, S.~{Bryan},
  H.-M. {Cho}, et~al.
\newblock {Development of Multi-chroic MKIDs for Next-Generation CMB
  Polarization Studies}.
\newblock {\em Journal of Low Temperature Physics}, 193:103--112, November
  2018.
\newblock \href {http://arxiv.org/abs/1711.02523} {\path{arXiv:1711.02523}},
  \href {http://dx.doi.org/10.1007/s10909-018-2032-y}
  {\path{doi:10.1007/s10909-018-2032-y}}.

\bibitem{Ji2014}
C.~{Ji}, A.~{Beyer}, S.~{Golwala}, and J.~{Sayers}.
\newblock {Design of antenna-coupled lumped-element titanium nitride KIDs for
  long-wavelength multi-band continuum imaging}.
\newblock In {\em Millimeter, Submillimeter, and Far-Infrared Detectors and
  Instrumentation for Astronomy VII}, volume 9153 of {\em \procspie}, page
  915321, July 2014.
\newblock \href {http://dx.doi.org/10.1117/12.2056777}
  {\path{doi:10.1117/12.2056777}}.

\bibitem{TIME-Pilot}
A.~T. {Crites}, J.~J. {Bock}, C.~M. {Bradford}, T.~C. {Chang}, A.~R. {Cooray},
  L.~{Duband}, et~al.
\newblock {The TIME-Pilot intensity mapping experiment}.
\newblock In {\em Millimeter, Submillimeter, and Far-Infrared Detectors and
  Instrumentation for Astronomy VII}, volume 9153 of {\em \procspie}, page
  91531W, August 2014.
\newblock \href {http://dx.doi.org/10.1117/12.2057207}
  {\path{doi:10.1117/12.2057207}}.

\bibitem{Bryan2016}
S.~{Bryan}, J.~{Aguirre}, G.~{Che}, S.~{Doyle}, D.~{Flanigan}, C.~{Groppi},
  et~al.
\newblock {WSPEC: A Waveguide Filter-Bank Focal Plane Array Spectrometer for
  Millimeter Wave Astronomy and Cosmology}.
\newblock {\em Journal of Low Temperature Physics}, 184:114--122, July 2016.
\newblock \href {http://arxiv.org/abs/1509.04658} {\path{arXiv:1509.04658}},
  \href {http://dx.doi.org/10.1007/s10909-015-1396-5}
  {\path{doi:10.1007/s10909-015-1396-5}}.

\bibitem{Barrentine2016}
E.~M. {Barrentine}, G.~{Cataldo}, A.~D. {Brown}, N.~{Ehsan}, O.~{Noroozian},
  T.~R. {Stevenson}, et~al.
\newblock {Design and performance of a high resolution {$\mu$}-spec: an
  integrated sub-millimeter spectrometer}.
\newblock In {\em Millimeter, Submillimeter, and Far-Infrared Detectors and
  Instrumentation for Astronomy VIII}, volume 9914 of {\em \procspie}, page
  99143O, July 2016.
\newblock \href {http://dx.doi.org/10.1117/12.2234462}
  {\path{doi:10.1117/12.2234462}}.

\bibitem{Endo2012}
A.~{Endo}, P.~{Werf}, R.~M.~J. {Janssen}, P.~J. {Visser}, T.~M. {Klapwijk},
  J.~J.~A. {Baselmans}, et~al.
\newblock {Design of an Integrated Filterbank for DESHIMA: On-Chip
  Submillimeter Imaging Spectrograph Based on Superconducting Resonators}.
\newblock {\em Journal of Low Temperature Physics}, 167:341--346, May 2012.
\newblock \href {http://arxiv.org/abs/1107.3333} {\path{arXiv:1107.3333}},
  \href {http://dx.doi.org/10.1007/s10909-012-0502-1}
  {\path{doi:10.1007/s10909-012-0502-1}}.

\bibitem{Parshley18}
S.~C. {Parshley}, J.~{Kronshage}, J.~{Blair}, T.~{Herter}, M.~{Nolta}, G.~J.
  {Stacey}, et~al.
\newblock {CCAT-prime: a novel telescope for sub-millimeter astronomy}.
\newblock In {\em Ground-based and Airborne Telescopes VII}, volume 10700 of
  {\em Society of Photo-Optical Instrumentation Engineers (SPIE) Conference
  Series}, page 107005X, July 2018.
\newblock \href {http://arxiv.org/abs/1807.06675} {\path{arXiv:1807.06675}},
  \href {http://dx.doi.org/10.1117/12.2314046} {\path{doi:10.1117/12.2314046}}.

\bibitem{Kim18SPT}
J.~{Kim}, D.~P. {Marrone}, C.~{Beaudoin}, J.~E. {Carlstrom}, S.~S. {Doeleman},
  T.~W. {Folkers}, et~al.
\newblock {A VLBI receiving system for the South Pole Telescope}.
\newblock In {\em Millimeter, Submillimeter, and Far-Infrared Detectors and
  Instrumentation for Astronomy IX}, volume 10708 of {\em Society of
  Photo-Optical Instrumentation Engineers (SPIE) Conference Series}, page
  107082S, July 2018.
\newblock \href {http://arxiv.org/abs/1805.09346} {\path{arXiv:1805.09346}},
  \href {http://dx.doi.org/10.1117/12.2301005} {\path{doi:10.1117/12.2301005}}.

\bibitem{Padin14}
S.~{Padin}.
\newblock {Inexpensive mount for a large millimeter-wavelength telescope}.
\newblock {\em \ao}, 53:4431--4439, July 2014.
\newblock \href {http://dx.doi.org/10.1364/AO.53.004431}
  {\path{doi:10.1364/AO.53.004431}}.

\bibitem{Bertoldi18}
F.~{Bertoldi}.
\newblock {The Atacama Large Aperture Submm/mm Telescope (AtLAST) Project}.
\newblock In {\em Atacama Large-Aperture Submm/mm Telescope (AtLAST)}, page~3,
  January 2018.
\newblock \href {http://dx.doi.org/10.5281/zenodo.1158842}
  {\path{doi:10.5281/zenodo.1158842}}.

\bibitem{deBreuck18}
Carlos De~Breuck.
\newblock Site considerations for atlast, January 2018.
\newblock URL: \url{https://doi.org/10.5281/zenodo.1158848}, \href
  {http://dx.doi.org/10.5281/zenodo.1158848}
  {\path{doi:10.5281/zenodo.1158848}}.

\bibitem{Mroczkowski18}
T.~{Mroczkowski} and O.~{Noroozian}.
\newblock {AtLAST Instrumentation Considerations and Overview}.
\newblock In {\em Atacama Large-Aperture Submm/mm Telescope (AtLAST)}, page~26,
  January 2018.
\newblock \href {http://dx.doi.org/10.5281/zenodo.1159053}
  {\path{doi:10.5281/zenodo.1159053}}.

\bibitem{pico18}
B.~M. {Sutin}, M.~{Alvarez}, N.~{Battaglia}, J.~{Bock}, M.~{Bonato},
  J.~{Borrill}, et~al.
\newblock {PICO - the probe of inflation and cosmic origins}.
\newblock In {\em Space Telescopes and Instrumentation 2018: Optical, Infrared,
  and Millimeter Wave}, volume 10698 of {\em Society of Photo-Optical
  Instrumentation Engineers (SPIE) Conference Series}, page 106984F, July 2018.
\newblock \href {http://arxiv.org/abs/1808.01368} {\path{arXiv:1808.01368}},
  \href {http://dx.doi.org/10.1117/12.2311326} {\path{doi:10.1117/12.2311326}}.

\bibitem{CMB-Bharat}
Tarun~Souradeep et~al.
\newblock {CMB Bharat: Assessing the prospects for frontier CMB space
  experiments from India}.
\newblock \url{http://cmb-bharat.in/}, 2018.
\newblock [Online].

\bibitem{PRISM2013}
{PRISM Collaboration}, P.~{Andre}, C.~{Baccigalupi}, D.~{Barbosa},
  J.~{Bartlett}, N.~{Bartolo}, et~al.
\newblock {PRISM (Polarized Radiation Imaging and Spectroscopy Mission): A
  White Paper on the Ultimate Polarimetric Spectro-Imaging of the Microwave and
  Far-Infrared Sky}.
\newblock {\em arXiv e-prints}, June 2013.
\newblock \href {http://arxiv.org/abs/1306.2259} {\path{arXiv:1306.2259}}.

\bibitem{COREmission}
J.~{Delabrouille}, P.~{de Bernardis}, F.~R. {Bouchet}, A.~{Ach{\'u}carro},
  P.~A.~R. {Ade}, R.~{Allison}, et~al.
\newblock {Exploring cosmic origins with CORE: Survey requirements and mission
  design}.
\newblock {\em \jcap}, 4:014, April 2018.
\newblock \href {http://arxiv.org/abs/1706.04516} {\path{arXiv:1706.04516}},
  \href {http://dx.doi.org/10.1088/1475-7516/2018/04/014}
  {\path{doi:10.1088/1475-7516/2018/04/014}}.

\bibitem{Spica16}
H.~{Kaneda}, D.~{Ishihara}, S.~{Oyabu}, M.~{Yamagishi}, T.~{Wada}, M.~{Kawada},
  et~al.
\newblock {SPICA Mid-infrared Instrument (SMI): technical concepts and
  scientific capabilities}.
\newblock In {\em Space Telescopes and Instrumentation 2016: Optical, Infrared,
  and Millimeter Wave}, volume 9904 of {\em \procspie}, page 99042I, July 2016.
\newblock \href {http://dx.doi.org/10.1117/12.2232442}
  {\path{doi:10.1117/12.2232442}}.

\end{thebibliography}

\end{document}